\documentclass[manuscript,screen,nonacm]{acmart}
\AtBeginDocument{%
  \providecommand\BibTeX{{%
    \normalfont B\kern-0.5em{\scshape i\kern-0.25em b}\kern-0.8em\TeX}}}

\setcopyright{acmcopyright}
\copyrightyear{2023}
\acmYear{2023}
\acmDOI{XXXXXXX.XXXXXXX}

\acmConference[FAccT]{June 12--15,
  2023}{Chicago, USA}
\begin{document}

\title{Manifestations of Xenophobia in AI Systems}

\author{Nenad Tomasev}
\email{nenadt@deepmind.com}
\orcid{0000-0003-1624-0220}
\affiliation{%
  \institution{DeepMind}
  \city{London}
  \country{United Kingdom}
}

\author{Jonathan Leader Maynard}
\email{jonathan.leader_maynard@kcl.ac.uk}
\affiliation{%
  \institution{King's College London}
  \city{London}
  \country{United Kingdom}
}

\author{Iason Gabriel}
\email{iason@deepmind.com}
\orcid{0000-0002-4412-1686}
\affiliation{%
  \institution{DeepMind}
  \city{London}
  \country{United Kingdom}
}



\begin{abstract}
Xenophobia is one of the key drivers of marginalisation, discrimination, and conflict, yet many prominent machine learning (ML) fairness frameworks fail to comprehensively measure or mitigate the resulting xenophobic harms. Here we aim to bridge this conceptual gap and help facilitate safe and ethical design of artificial intelligence (AI) solutions. We ground our analysis of the impact of xenophobia by first identifying distinct types of xenophobic harms, and then applying this framework across a number of prominent AI application domains, reviewing the potential interplay between AI and xenophobia on social media and recommendation systems, healthcare, immigration, employment, as well as biases in large pre-trained models. These help inform our recommendations towards an inclusive, xenophilic design of future AI systems.
\end{abstract}



\keywords{algorithmic fairness, ethics, machine learning, artificial intelligence, marginalised groups, xenophobia}


\received{10 May 2023}
\received[revised]{}
\received[accepted]{}

\maketitle

\section{Introduction}
\label{sec:intro}

The increasing scope of the field of AI ethics -- and corresponding analyses of algorithmic fairness -- reflect the ubiquitous nature of AI deployment across many spheres of daily life. A major consequence of this is an increasing need to operationalize standards of algorithmic fairness widely, in order to minimise the risks involved with the use of technology and ensure an equitable distribution of its benefits. To date, much attention has focused on issues of fairness and discrimination associated with the legally protected categories of race and gender. However, research has shown that these frameworks tend to be be ill-suited for unobserved or less frequently-recorded characteristics such as sexual orientation and gender identity~\citep{queerfairness}. Building upon this insight, the present paper focuses on a prominent and widespread form of discrimination which has received comparatively little attention in ML fairness research -- xenophobia, which is discrimination directed at the \emph{other} or at those who are perceived to be \emph{foreign}. 

Xenophobia shapes the political landscape of the modern world in a number of ways~\citep{bowman2021xenophobia, arredondo2018latinx, kopyciok2021left, yakushko2018modern, akinola2018political, chenzi2021fake} -- as can be seen in the recent growth of authoritarian populist movements and  anti-immigration political platforms~\citep{populismxeno, altright}. Moreover, the Covid-19 pandemic led to a sharp rise in xenophobia~\citep{xenocovid}, and in particular to a rise in anti-Asian sentiment in the United States~\citep{ijerph17197032}. There is, consequently, a moral imperative to develop a better understanding of the role AI technology plays in amplifying or mitigating xenophobia in society. Indeed, as David Haekwon Kim and Ronald R. Sundstrom observe, by ignoring xenophobic discrimination – as distinct from racial, sexual or gender discrimination – we risk neglecting specific patterns of disadvantage and harm associated with conceptions of ”foreignness”. This omission then leads to a situation in which “a normative loophole” is created, one that holds that \emph{“when thinking about social, national justice we need not care about foreign distant others.”}~\citep{Kim2014-KIMXAR}

From a technical standpoint, existing bias-mitigation strategies in the domain of ML fariness tend to focus on ensuring equitable outcomes for legally protected groups. However, efforts to combat xenophobia in AI need to go further, as many groups that are deemed to be foreign lack such legal protection. An additional challenge lies in the need to understand the decisions people make when distinguishing between \emph{us} and \emph{them}, between those who familiar and those who are not. Ultimately, political institutions often help define and impose such categorizations, with the history of the US census providing an example of how changes in data collection may reflect and entrench changing notions of identity over time. 

In the context of AI research, the detection of “hate speech” or “dangerous speech” has served as a focal point for efforts to address xenophobia, fuelling the development of content moderation systems to tackle the spread of hateful and dangerous speech in social media~\citep{hatespeech1, hatespeech2, hatespeech3, hatespeech4}. Dangerous speech is an element of a path of escalation that may lead to conflict~\citep{bowman2021xenophobia} and  even mass atrocities~\citep{dangerousspeech, dangerousspeechmyanmar} -- while early detection of dangerous speech can provide opportunities for timely deescalation. However, accurate hate speech detection is not without challenges, as early systems have been shown to incorporate racial bias~\citep{hatespeechracialbias}, as well as over-reporting queer speech as inappropriate~\citep{gomes2019drag}. We believe that the existing focus on social media platforms needs to be complemented by deeper analysis of xenophobia, spanning the full range of AI use cases. Such analysis needs account for, and be rooted in, recognition of structural barriers facing foreign nationals and immigrants. 

To the best of our knowledge, this paper is the first comprehensive review of the interplay between xenophobia and AI. In total we look at its impact across five domains: social media, healthcare, immigration, employment, as well as via large pre-trained models. We proceed as follows. First, in Section~\ref{sec:xenobackground}, we define xenophobia, distinguishing it from other forms of discriminatory attitudes and practices such as racism and sexism, and specifying its conceptual relevance to AI systems. We then, in Section~\ref{sec:considerations} discuss how xenophobia may manifest across a number of AI application domains. We proceed, in Section~\ref{sec:xenophilic} to make a moral argument towards a design of inclusive, xenophilic systems, coupled with a set of technical considerations and recommendations informed by the use cases reviewed in Section~\ref{sec:considerations}. We conclude by outlining key issues for future research.

\section{On Xenophobia}
\label{sec:xenobackground}

Xenophobia is commonly understood as a kind of hostility or prejudice directed towards foreigners, immigrants or, more broadly, those construed as “others”. It can manifest as a fear, dislike or hate towards people who are perceived to be different, including their culture and customs, and has been associated with a range of social and psychological roots, \citep{xenocultural, sanchez2015xenophobia, xenoracismexpl, xenoroots} including misassociations, stereotyping and cognitive bias~\citep{xenologic}, projections onto the other, as well as active processes of othering groups of people~\citep{xenoracismexpl}. Moving beyond the cognitivist conception of xenophobia -- which focuses primarily upon mental states and attitudes -- it is important to recognize that xenophobia also has institutional and structural expressions. In these cases discriminatory practices or outcomes are sustained by rules, procedures, and prior distributions of material and symbolic resources which target specific groups of people. Xenophobic discrimination may involve both phenomenon, typically involving a mix of attitudinal and institutional effects~\citep{Kim2014-KIMXAR}. 

This conceptual core, however, leaves two important questions unanswered: the first concerns the distinction between xenophobia and other forms of intergroup prejudice, notably racism (the Distinctiveness Problem); and second centers upon defining the threshold of what counts as xenophobia, given the ubiquitous differences in how rights and responsibilities are allocated to citizens and non-citizens in contemporary nation-states (the Threshold Problem).

Much scholarship on discrimination and disadvantage ignores the first problem, by adopting, as Kim and Sundstrom observe, “the general assumption, alive in popular culture and some segments of the academy, that racism has subsumed nativism and xenophobia”~\citep{Kim2014-KIMXAR}. This, they point out, is problematic, since it “merges racism, xenophobia, and nativism into one hyper-concept of prejudice and exclusion” and ignores “important distinctions to be made here between prejudice against racial outsiders, civic outsiders, and the pursuit of chauvinistic ethics and racial group-interests based on claims of indigenousness.”~\citep{Kim2014-KIMXAR} While the history of xenophobia can not be decoupled from the artifacts of the colonial past and racism, xenophobia is conceptually distinct from racism, and the strategies needed to mitigate racism and xenophobia may differ~\citep{xenoracism}.

We adopt a modified version of Kim’s and Sundstrom’s solution to this problem: namely, orientating our understanding of xenophobia around the notion of “civic ostracism”. Specifically, we distinguish xenophobia from other forms of prejudice and discrimination by defining it as a \textit{discriminatory orientation (whether of individuals or institutions) that penalises individuals on the basis of their perceived foreignness, understood here as the purported lack of full membership of the civic community}. Such non-membership may correlate with actual legal status, but it need not do so: individuals who are in fact citizens but who are prejudicially deemed foreign (perhaps because they were originally born outside the civic community, or simply because of assumptions associated with accent, name, practices, beliefs or race) may still be subject to xenophobia in this conceptualisation.

Kim and Sundstrom do not address, however, the second problem – which goes persistently unaddressed in most scholarship on xenophobia. By their very nature, nation-states distinguish between citizens and non-citizens, and in doing so, allocate rights, responsibilities and resources differentially between these two categories. Adherents to “cosmopolitan” ethics may lament this institutional reality of contemporary politics~\citep{caney2005justice, pogge2002cosmopolitanism, benhabib2008another},  yet the ethical significance of civic national groups that differentiate between citizens and non-citizens has cogent defenders, and the abandonment of civic nations is extremely unlikely for the foreseeable future. A framework which deems any differential outcomes between citizens and non-citizens to be xenophobic is, as such, unlikely to be accepted as offering relevant ethical guidance for AI systems. At the same time, differential outcomes for non-citizens can indeed arise from xenophobia,  and may extend up to persecution, ethnic cleansing, and genocide~\citep{mann2005dark, straus2015making, brubaker1998ethnic}. The recognition of such systemic and structural xenophobia must play a central role in understanding the potential xenophobic impact of AI systems.

Our proposed solution to these ambiguities, when assessing xenophobia in technical systems, is to distinguish between two different ways of operationalising the concept of xenophobia. In general, we think designers of AI systems will be interested, in the first instance, with what we term \textit{immanent assessments} of xenophobia, by which we mean the determination of whether an AI system creates disadvantages on “foreigners” that are discriminatory according to the societal standards. Put another way, if AI systems generate disadvantages that go beyond the officially sanctioned differences in the rights, responsibilities and resources that separate citizens from non-citizens, the system is immanently xenophobic, violating the standards formally adhered to by its user. Yet those standards may, themselves, be subject to ethical critique as part of a more \textit{transcendent assessment} of xenophobia – i.e. they violate ethical principles which, even if the actor or organization does not formally recognise or accept them, ought to be upheld. While we think transcendent assessments of xenophobia are crucial to ethical debate in global affairs, much of the time AI-systems may be rendered substantially fairer simply by ensuring that immanently understood xenophobia is avoided.

Whether we are assessing xenophobia immanently or transcendentally, we identify three main kinds of harms that may flow from a xenophobic AI system. First, a xenophobic system may result in \textit{discriminatory material disadvantages} – allocating material resources in ways that systematically and unfairly penalise certain individuals on the basis of their actual or presumed foreignness. Second, a xenophobic system may deny individuals \textit{proper ethical recognition}, by formally or informally engaging with them in ways that communicate effective civic ostracization. Third, a xenophobic system may \textit{restrict the effective exercise of individuals’ rights}, for example, by communicating information about them that biases people from engaging with them in a fair way, or unfairly allocating access to forms of legal action, or giving unjustified prominence to certain groups over others in platforms for expressing speech. Such restriction of rights could include exposing individuals to outright violence or physical harm if, for example, law enforcement officers rely on systems that contain xenophobic bias. Moreover, while these harms may overlap with discrimination based on other attributes, such as race and gender, they are not subsumed by these categories. More obviously, for example, xenophobic discrimination may be directed at immigrant and refugee populations, even when such groups are not clearly perceived of as ‘racially’ distinct.
\section{Practical considerations}
\label{sec:considerations}

We now focus on several important AI application domains to establish different ways in which xenophobic discrimination may potentially manifest, and where countervailing measures may need to be adopted. Since our aim here is to draw attention to an aspect of fairness – xenophobia – which has thus far rarely been addressed by AI researchers, and for which there is very little known testing, the evidence we draw upon is necessarily suggestive and somewhat indirect, drawing heavily upon analogies and examples from the wider sociotechnical literature. Our contention is simply that such evidence gives sufficient reason to think that xenophobia is likely to to manifest in AI systems, and that there should be more urgency in developing frameworks for evaluating potential xenophobic bias and downstream harms.

\subsection{Social media}

Social media may amplify xenophobia~\citep{daniels2018algorithmic}, especially in communities with a certain amount of pre-existent xenophobic sentiment~\citep{bursztyn2019social, yamaguchi2013xenophobia}. Low barriers to entry, reliance on user-generated content, difficulties in moderation, and the emergence of filter bubbles~\citep{baldauf2019hate, khosravinik2017right, bruns2019filter, zuiderveen2016should} present difficulties that may lead to the easier spread of fake news~\citep{albright2017welcome, schafer2018online} and potentially dangerous and hateful speech \citep{dangerousspeech} directed at those who are deemed foreign. At the same time, these platforms could also potentially provide mechanisms to enable more inclusive discourse, expressions of support and positive views towards marginalised groups, serving as a medium for communicating information relevant to the effective exercise of individual rights. This speaks to a dual role social media platforms occupy in shaping culture: both as a source of greater cross-cultural understanding, and also as a mechanism that can entrench social divisions, sometimes leading to abuse that reflects wider structural and cultural violence~\citep{haunschild2022cultural}. In particular, it has been shown that there is an important link between social media and hate crimes, as the prevalence of anti-refugee sentiment on social media platforms is a strong predictor of crimes against refugees across municipalities~\citep{muller2021fanning}.

The dynamics of discourse and content sharing on social media are ultimately influenced both by the societal context as well as the design of the platform, its implemented algorithmic solutions, and the enforcement of content moderation policies. Carefully designed platforms may present safe spaces for information sharing and help facilitate collective immigrant action~\citep{zepeda2016weapons}. In general, given an ever-increasing proportion of time spent on social media, its overall impact on the amount and the intensity of xenophobia in society can be profound~\citep{postill2018populism}.

\subsubsection{Risks}

All three types of xenophobic harms that we consider in this study may occur in the context of social media applications. To begin with, discrepancies across personalised social media feeds often incorporate harmful stereotypes and make implicit assumptions about people that are problematic from the standpoint of representation. Indirectly, these cultural tropes may shape the public opinion in ways that result in xenophobic discrimination offline, exacerbating the propensity to exclude individuals based on their perceived “foreignness”. Indeed, xenophobic harms pertaining to civic ostracism may follow, for example, from online public shaming~\citep{aitchison2021against} and amplified hate speech aimed at marginalised communities. Finally, indirect xenophobic harm may result indirectly from the shaping of public opinion, as well as directly by silencing minority views~\citep{oliva2021fighting, haimson2021disproportionate} and weaponising content moderation against foreign individuals, resulting in their effective exclusion from public forums and an inability to influence future decisions. To address these concerns, greater transparency and contestability need to be incorporated in content moderation system design~\citep{vaccaro2020end}, to ensure fair processes and help mitigate the consequences of potential algorithmic biases.

Machine learning plays an important role in emerging patterns of social media content consumption. Personalized news feeds may give rise to feedback loops~\citep{jiang2019degenerate}, and amplify pre-existing opinions and biases, resulting in \emph{autopropaganda}~\citep{whittaker2021recommender} Moreover, repeated exposure to content that others out-groups and reinforces harmful stereotypes may lead to radicalisation~\citep{alfano2018technological, o2015down, ribeiro2020auditing}, suggesting that a careful redesign of content consumption pathways is needed~\citep{fabbri2022rewiring}. Deepfakes~\citep{westerlund2019emergence} present a recent manifestations of the risk of AI-enhanced spread of misinformation, either by malicious individuals and organisations, or as a part of state propaganda~\citep{pavlikova2021propaganda}. Emergence of deepfake detection technology~\citep{guera2018deepfake, dolhansky2019deepfake, zhao2021multi, wang2021m2tr, fagni2021tweepfake} may help address this challenge, at least until the deepfake technology becomes advanced enough to be indistinguishable from authentic images, video, and audio streams. Adversarial robustness of deepfake detection models plays a crucial role in this context~\citep{neekhara2021adversarial}, given the known vulnerabilities of neural computer vision systems to adversarial perturbations. Yet, technological mitigations for deepfake dissemination also need to be accompanied by an investment in media literacy~\citep{hwang2021effects} and user awareness. Disinformation and deception have been shown to play a key role in forming xenophobic narratives around otherness, aiming to associate foreigners with acts of violence, over-reliance of financial aid, and illegal immigration~\citep{gamir2021multimodal}. 

\subsubsection{Promise}

Advances in natural language processing may help develop more reliable hate speech detection~\citep{badjatiya2017deep, aluru2020deep, gamback2017using, rizos2019augment, sutejo2018indonesia, plaza2020detecting, khan2020hateclassify, alatawi2021detecting} and sentiment prediction systems~\citep{jin2020multi, seo2020comparative}, as well as improved automated topic analysis~\citep{gui2019neural, WANG2019102098} in social media posts, across different modalities~\citep{cmtopicmodel}. This improved monitoring may also help facilitate targeted interventions (e.g. counter-speech~\citep{sonntag2019social, hatecounter}) to mitigate xenophobic sentiment, safeguard marginalised communities, and minimise the risk of escalation towards potential violent conflict. Complementary efforts in semi-automatic detection of fake news~\citep{popat2018declare, wang2020weak, kaliyar2021fakebert}, as well as bots and troll accounts~\citep{luceri2020detecting} are imperative in establishing safe online spaces. Human-in-the-loop~\citep{strickland2018ai, demartini2020human} filtering processes and explainability in AI-driven content moderation may be necessary to address biases in language systems.

Advances in diversification of recommender systems~\citep{kunaver2017diversity, helberger2018exposure, mansoury2020fairmatch} and improvements in user fairness~\citep{leonhardt2018user, nandy2022achieving} may help avoid the formation of filter bubbles and echo chambers and empower communities to confront and challenge xenophobic narratives. Content diversity needs to be accompanied by additional counter-measures to be effective~\citep{stray2021designing} and more research is needed to identify the best contextual approach to use. In particular, neural machine translation~\citep{bahdanau2014neural, chen2018best, aharoni2019massively, pan2021contrastive} approaches may prove to be an invaluable tool in breaking down barriers in communication across groups, increasing exposure to contrasting views, and help establish multicultural online spaces. To achieve this, it is important to improve the accessibility of natural language understanding, in particular in terms of improving the performance of existing systems in low-resource languages~\citep{gu2018meta, gu2018universal, karakanta2018neural, sennrich2019revisiting, siddhant2020leveraging}, through participatory action and community engagement. Finally, ML also offers opportunities for developing tools for monitoring large-scale shifts in social behaviour~\citep{papakyriakopoulos2020political, patti2017ethical}, which may help further the development of theories that aim to understand the root causes behind an increase in xenophobic sentiment. These theories may be driven by observations in retrospective traces of social media usage~\citep{frias2019hate, wahlstrom2021dynamics}, or in simulation~\citep{yao2021measuring, lucherini2021t}, with controllable parameters so as to isolate individual effects.
\subsection{Immigration}

Perhaps the most obvious context in which AI systems may exacerbate or mitigate xenophobic harms is in the management of migration flows. The complexity of immigration policies reflects a tension between internal security and the liberal frame of humanitarianism~\citep{lavenex2001migration}, generating a need for involvement of stakeholders outside of established political processes. Migration between culturally distinct countries and regions is often accompanied by a growing resentment of immigrants and refugees~\citep{hadvzic2020european}. Meanwhile, denialism and complicity with populist narratives hamper the implementation of remedial measures -- including the development of a coordinated strategy, spanning international, national and regional efforts, which is needed given the magnitude of the problem~\citep{xenophobiamigrationdev}. Moreover, states have already been quick to employ new digital and AI technologies in an effort to more tightly and effectively control migration~\citep{nalbandian2022eye}. As an official from the South African National Defence Force observes: “The European Union is trying to fight the issue [of immigration] with data… In the modern digital world, you cannot exist without a digital presence. Why is it not feasible to control all of this immigration by tracking the behavior of this data, within the database?” \citep{longo2017politics}. While digital technology may promise significant gains in efficiency in the management of human movement across borders, this opens a vast realm of potential bias and harm created by AI systems that restrict rights of movement based on assumptions of risk or threat associated with certain data profiles. The nature of problems faced by immigrants is necessarily intersectional, as immigrants from marginalised communities face additional barriers when attempting to integrate into society~\citep{BHAGAT2018155}, and may employ identity concealment tactics to render themselves invisible in the face of xenophobic attitudes~\citep{makoni2020metalinguistic, tewolde2021passing}.

\subsubsection{Risks}

Numerous technological solutions incorporating AI have been proposed for border control and migrant assessment. The collection and utilisation of data from refugees, asylum–seekers and migrants -- including by obtaining access to their private social media accounts~\citep{andreassen2021social} -- poses serious risk of misuse\citep{algorithmichumanitarianism}. AI systems have also been proposed for automatic assessment of immigration forms~\citep{egovimmi}, and the deployment of facial recognition technology is also a feature of this context~\citep{8489113}. This risks perpetuating the historical, national, ethnic, and racial stereotypes. Indeed, many governments have been quite explicit in their efforts to define migrants in terms of an algorithmic computation of 'risks' -- with migration processing dictated by the resulting assessment of risk levels. For example, officials in the Australian Department of Immigration and Citizenship have publicly defended the employment of an electronic travel card system on the grounds that it enables a “risk-tiered approach to identity…where we have a range of concentric circles around Australia and we are pushing [the border] further and further out, [and] are doing more and more checks before the person hits our shores” \citep{longo2017politics}. 

While the management of migration risks~\citep{illegalimmi} may be legitimate in principle, such an approach creates enormous potential for harmful and unfair outcomes to individuals based on associative assumptions about risk that are built into AI assessment systems. These harms can potentially manifest in different walks of life - for example, systems for the assessment of the likelihood of the H1B visa approval in the US immigration process~\citep{h1bimmi} can bias potential employers against certain immigrant groups. There is also an important distinction to be made here between immanent and transcendental standards in assessing xenophobic harms in immigration, where some practices may be legally permissible, yet palpably unjust. This raises important questions about migrants' human rights in face of the use of AI technology in immigration enforcement~\citep{giannakou2021migrants}.

\subsubsection{Promise}

If xenophobic bias can be avoided, AI systems offer the potential for fairer, faster and more efficient management of migration flows with benefits not only for governments, in encouraging welcome forms of migration, but also for migrants in avoiding complex, lengthy and expensive migration processes. Indeed, AI systems have been proposed as a mechanism for encouraging states to more effectively and expansively house refugee populations, by speedily “matching” government and individual preferences and needs in the context of large refugee flows~\citep{refugee_match}. Nonetheless, improved auditing of such systems is required to reduce the risk of unjustly excluding people from safe refuge. Advances in modelling and understanding trends in human migration~\citep{migmodel} can also help with anticipatory measures and planning towards building sustainable and inclusive communities. Modelling refugee flows~\citep{mead2020creating} may prove helpful in expediting a humanitarian response.

\subsection{Healthcare}

Xenophobia has recently been identified as an important determinant of health~\citep{suleman2018xenophobia}, with a multitude of adverse effects being reported for both individual and community-based health metrics. Moreover, policies that restrict the range of health services available to foreigners sometimes exceed the costs they purport to be saving, creating a hostile environment both for patients and migrant workers alike ~\citep{shahvisi2019austerity}. Exposure to repeated xenophobic prejudice in a healthcare context may erode trust and adversely affect care-seeking patient behaviours~\citep{earnshaw2019medical}. 

Rather than being an exclusive artefact of misguided public health policy, medical xenophobia can manifest via negative attitudes and practices by health workers, and has been shown to be deeply entrenched in numerous public health systems~\citep{crush2014medical, loganathan2019breaking}. Discriminatory practices are further amplified in case of particularly vulnerable migrant populations, like refugees~\citep{zihindula2017lived, munyaneza2019medical} and undocumented migrants~\citep{richter2015medical}. Moreover, even prior to engaging with health services, xenophobia and racism may impact health via early exposure to adverse childhood experiences~\citep{nguyen2021deconstructing}. In Adja et al. ~\citep{adja2020social}, the authors point out that the World Health Organization (WHO) defines health as “a state of complete physical, mental and social well-being and not merely the absence of disease or infirmity“~\citep{world1946preamble} and that social well-being is often overlooked in conversations around health. A holistic approach for mitigating the effects of medical xenophobia must therefore incorporate notions of social well-being and account for the impact of the social determinants of health and the disparities faced by foreigners outside of the boundaries of direct medical care.

Medical xenophobia has also historically played an important role in shaping epidemiological perception related to the spread of infectious diseases. More recently, this could be seen in the overwhelmingly xenophobic response to the spread of COVID-19~\citep{reny2020xenophobia, le2020anti} and the resulting health disparities~\citep{hooper2020covid}, but it is by no means a recent phenomenon. As highlighted in~\citep{sarbu2014brief, galassi2020sinophobia}, the act of blaming foreigners for the spread of infectious disease could be readily seen in the accounts related to the devastating epidemic of venereal syphilis in Europe in the 15th and 16th century. The English, Germans and the Italians blamed the French for the epidemic, referring to syphilis as \emph{morbus Gallicus} (the “French disease”). In turn, the French accused the Germans, the Poles and the Neapolitans, the Poles blamed the Germans, the Russians blamed the Poles, etc~\citep{sarbu2014brief, galassi2020sinophobia}. Paleopathology can help alert us to past instances of medical xenophobia, raise public awareness and reduce future xenophobic discrimination in times of public health emergency. As articulated by Muscat et al. ~\citep{muscat2017public}: “Public health is not a neutral scientific activity, but a political activity informed by science. Public health has a moral mandate that complements its scientific mandate.“

\subsubsection{Risks}

Xenophobic medical AI harms include all three types of xenophobic harm that we consider in this study. Clinical AI systems developed from retrospective clinical data may encode and perpetuate harmful biases towards migrants and ethnic minorities, potentially leading to discriminatory material disadvantages - misdiagnosis, suboptimal treatment decisions and resource allocation, and worse health outcomes~\citep{suleman2018xenophobia}. Design choices in setting up the critical health data infrastructure play a role in this, in ways in which they encode identities and represent demographics~\citep{yi2022mutually}. AI systems that take shortcuts~\citep{degrave2021ai, alex_shortcuts}, through proxies of sensitive attributes, may further ostracize communities seen as foreign and perpetuate narratives linking foreignness to disease~\citep{yi2022mutually}. Exclusion of ethnic groups may potentially also follow from non-inclusive designs of systems that may not be suited for safe application beyond the majority demographics in the country of deployment, resulting in disproportionate rates of deferral to alternative pathways, increased waiting times, and a deep misunderstanding of patient needs. Finally, clinical AI systems may further alienate patients and erode trust, as well as potentially reinforce harmful xenophobic stereotypes in repeated interactions with systems designed to reason based on historical data.

\subsubsection{Promise}

There is potential for AI technologies in healthcare to play a transformative role in improving access to care and the quality of its delivery. Advances in deep learning have led to a number of promising early research prototypes for improved diagnostics in medical imaging~\citep{de2018clinically, mckinney2020international, baltruschat2019comparison}, diagnostics from wearables~\citep{ravi2016deep}, early detection of adverse events from electronic health records~\citep{tomasev2019clinically, tomavsev2021use, futoma2017learning}, and others. By way of illustration, on-device screening apps may help with earlier detection of health problems in marginalised groups, including migrants and foreign workers with limited access to health insurance and health services. For example, AI-based diabetic retinopathy detection from smartphone-based fundus photography was shown to have very high sensitivity~\citep{rajalakshmi2018automated} and would potentially be applicable for mass screening of at-risk populations. Similarly, advances in AI-based dermatology screening applications~\citep{gocceri2020impact} could yield similar benefit in the future. 

Yet, there are still practical obstacles for these types of systems to be safely deployed at scale~\citep{seneviratne2020bridging}. Participatory approaches play a key role in medicine~\citep{abma2010patient, angel2015challenges, falk2019barriers, schinkel2018perceptions}, where engaging with patient focus groups helps the clinicians align on the desired outcomes and improve the overall patient experience. Participatory approaches to developing AI solutions have similarly been identified as a key ingredient in safe and inclusive AI system design~\citep{Bondi_2021}, though their implementation is not without challenges~\citep{smith2021decolonizing}. The outcomes of such practices hinges on the inclusion of all marginalised communities and especially those that may be facing difficulties accessing the basic healthcare services, or facing language barriers~\citep{um2018application, nickell2019engaging, bradby2019undoing}.
\subsection{Employment}

AI systems are increasingly used in decision-making around employment~\citep{upadhyay2018applying, sanchez2020does}, in helping to identify, filter, assess and prioritise potential job candidates. Given that differential outcomes in employment translate to direct material harms, it is crucial to understand how xenophobia manifests in this sector overall, and how it may impact the development of AI-driven decision-support systems~\citep{raghavan2020mitigating}.

Unequal access to employment opportunities and unfair compensation are among the key drivers of social inequality as differences in socioeconomic status tend to have a knock-on effect on community health, mental health, access to housing and physical safety. Labour market attachment is also critical to the livelihood and identity formation of individuals and groups, especially when it comes to those that have been historically marginalised~\citep{teelucksingh2007working}. There are a number of different ways in which employment discrimination manifests for foreign workers~\citep{kosny2017employment} at different stages in the employment process, depending upon their citizenship and immigration status. Both citizens and non-citizens can suffer xenophobic harms in the employment process, although there are commonalities and differences in their respective experiences~\citep{bell2010immigrants}. Historically, the worst jobs, with the hardest working conditions and the least pay had been reserved for immigrants, leading to the subjugation of immigrant bodies and population control, fueled by the xenophobic sentiment -- which persists to this day~\citep{longhi2012immigrant}.

Immigrants play an important role in the labour market, due to their diverse skill sets, life experience, educational profiles, and background. This diversity necessitates an intersectional analysis of employment outcomes, as it has been shown that different immigrant groups experience different levels of discrimination. For example, a study in Switzerland identified highly competitive immigrant groups from neighboring countries as being subjected to most workplace incivility~\citep{krings2014selective}, presenting a skills paradox~\citep{dietz2015skill}. Such discrimination is not restricted to first-generation immigrants, rather being rooted in assumptions about heritage, names~\citep{midtboen2014invisible} or appearance~\citep{fibbi2006unemployment}. In France, both immigrants and their descendants face worse employment outcomes~\citep{duguet2010young, meurs2017role}. Meanwhile, in Sweden, replacing Swedish with Middle-Eastern-sounding names in CVs~\citep{carlsson2007evidence} was shown to reduce the rate of call-backs by a large margin, with the rate of discrimination being amplified by male recruiters in particular. While labour-market integration in Sweden remains straightforward for Western immigrants, immigrants from Africa, Asia and South America~\citep{grand2002permanent} face far greater obstacles, showcasing the negative impact of xenophobia. Similar trends have been observed in the United Kingdom~\citep{anderson2006fair}, where employers confessed to ethnic stereotyping in their job candidate preferences. Female job-seekers~\citep{jureidini2005migrant, dlamini2012negotiated} may experience further obstacles, necessitating an intersectional analysis.

Finally, efforts to assimilate successfully play key role in modulating xenophobic employment discrimination~\citep{kee1994native} and mitigating the resulting pay gap~\citep{nielsen2004qualifications}, though out-groups may instead experience further alienation due to xenophobic attitudes~\citep{mayadas1992integration}. Language barriers, cultural differences, lack of networks, and social capital~\citep{behtoui2010social} present challenges to securing employment~\citep{hakak2010barriers}. This lack of integration also affects foreign entrepreneurs, facing obstacles in securing capital for their businesses~\citep{teixeira2007immigrant}. A lack of elasticity contributes to the wage gap between the immigrant and native job seekers, and is reflected in monopsonistic discrimination against immigrants. Findings from a study in Germany suggest not only that the employers hold more power over immigrant workers as a result of search friction and lower flexibility, but also imply that employers profit from discriminating against immigrants~\citep{hirsch2015there}.

Existing legal frameworks are often insufficient in protecting the rights of foreign employees and addressing the problem of xenophobia in the workplace~\citep{handayani2021dysfunction, mubangizi2021xenophobia}. This insufficiency underscores the need for a holistic approach to ensuring fair employment outcomes. Technology should aim to complement and assist numerous organisational initiatives, collective action, legislative and legal measures, administrative measures, political and educational action and international standards~\citep{taran2004challenging}.

\subsubsection{Risks}

Predicting prospective employee performance from retrospective data~\citep{mahmoud2019performance} carries the risk of perpetuating historical discrimination against out-groups~\citep{kim2018big, mujtaba2019ethical}, under the veneer of alleged objectivity. Such issues have already been identified in a number of deployed AI systems~\citep{dastin2018amazon, hsu2020can}. Discriminatory outcomes may arise even in absence of direct information on ethnic and national background, via proxy features~\citep{prince2019proxy}. Additionally, there is a pressing need for establishing firm guidelines and policies governing the fair usage of employment decision support tech~\citep{kim2021artificial}. Xenophobic discrimination poses unique challenges in this context, as natives and immigrant workers are often not granted the same rights and protections. This is in part due to the specific requirements for work permits in terms of skills and training, corporate sponsorship, minimum compensation etc. These requirements may differ between immigrant workers of different national and ethnic backgrounds. Polices aimed against xenophobic discrimination in employment, and the fairness on AI systems in employment, need to account for such legal distinctions. The overall impact of AI on the job market spans beyond its immediate utilisation in employment decisions, and should be seen in relation to how technology is reshaping the future of work -- including job quality -- across industries and regions. 

\subsubsection{Promise}

Assistive technologies have the potential to improve the overall candidate experience as well as help overcome human biases and help improve diversity and inclusion~\citep{daugherty2019using}. ML models may also be seen as a powerful lens into existing hiring practices~\citep{van2020hiring}, helping audit~\citep{liem2018psychology} and improve such processes. Recently developed frameworks for ML explainability~\citep{bhatt2020explainable} and fairness~\citep{cabrera2019fairvis} may be utilised towards this end. Simulations may also prove helpful for policy and process design~\citep{hu2017fairness, bower2017fair, hu2018short, candidatescreeningfairness, schumann2020we}. Based on the prevalence of xenophobic discrimination in historical hiring practice, as well as language technologies that underpin employment AI systems, we believe that there is a pressing need for improved transparency and explicit checks for fairness in decision recommendations for candidates of different national and ethnic background, as well as immigration status. If employment AI tech is ever to deliver on its promise, it needs to demonstrate a commitment to fight deeply ingrained manifestations of xenophobia in the workplace.
\subsection{Stereotypes in Large Pretrained Models}

Recent trends in AI development involve systems that are increasingly general in the scope of their application. Large language models (LLM)~\citep{lang_pretrained, chowdhery2022palm}, for example, can be fine-tuned~\citep{lang_finetuning} and adapted to a range of applications and use cases, becoming components of even larger and more complex AI systems. These systems are sometimes referred to as ''foundation models''~\citep{bommasani2021opportunities}, or ''base models'', in recognition of their emerging role as building blocks~\citep{foundationcomposed} in such systems. The development of such models usually involves large-scale pre-training on increasingly large datasets which are increasingly hard to curate. This poses unique safety and ethics concerns, due to a difficulty of anticipating the totality of harmful biases and stereotypes, as well as the variety of use cases stemming from their wide applicability~\citep{ganguli2022predictability}. Large language models may exhibit varying degrees of positive or negative sentiment across national~\citep{rae2021scaling, venkit2023nationality}, societal~\citep{khandelwal2023casteist} and religious groups~\citep{abid2021persistent, abid2021large}, inheriting features of historical and contemporary discourse. Text-to-image models have recently been shown to amplify demographic stereotypes at scale~\citep{large_models_text_image}. Multimodal systems~\citep{jaegle2021perceiver} introduce another layer of complexity, with cross-modal stereotypes~\citep{multimodaldataissues} becoming detectable when considering modalities in relation to each other. Any biases in base models have the potential to emerge across their downstream applications, making it especially important to evaluate such models and their underlying training data prior to a wider release.

As an illustrative example of the types of xenophobic bias that may creep into large pretrained models, here we consider the case of media and arts. Indeed, narratives shaping the very formation of national and group identities, the \emph{us} and \emph{them}, are deeply embedded in media as societal cultural artifacts, either altering~\citep{igartua2017enhancing} and subverting or reinforcing and perpetuating pre-existing power dynamics within and across societies. Stereotyping and projection are central to the act of \emph{othering}, as the \emph{other} get associated with unwanted, negative traits. Undertaking a critical re-examination of how xenophobia is expressed and disseminated in media and arts is critical in understanding the risks of inadvertently incorporating xenophobic stereotypes in AI systems developed on broad multimedia data at scale. Recent advances in AI creativity and art co-creation~\citep{dalle, imagen, ai_fashion} raise interesting questions around the role of AI-generated art and the cultural stereotypes potentially contained within~\citep{jalal2021fairness, dalle_eval}.

In case of literature, xenophobia manifests not only in terms of harmful narratives leading to  discrimination~\citep{taras2009transnational} and violence~\citep{minga2015xenophobia}, but also via ethnolinguistic purity, centering only the works written by the favoured in-group as “true” national cultural heritage, whilst erasing contributions written in minority languages~\citep{kamusella2021xenophobia}. Xenophobia may also manifest in the lack of availability and literary translation of foreign works~\citep{dickens2002literary}, thereby sheltering the local xenophobic narratives from foreign critique. Portrayals of out-groups in film provide ample evidence for deeply rooted xenophobia, involving racial stereotyping~\citep{yuen2016reel}, unfavourable portrayals of Arabs~\citep{shaheen2003reel}, antisocial portrayal of LatinX people~\citep{berg2002latino}, stereotyping queer people as villains~\citep{long2021queer, brown2021hook}, etc. These works reinforce notions of superiority for the majority, while simultaneously promoting unbefitting images and traits with respect to out-groups~\citep{ullah2017cultural, benshoff2021america}. The pervasiveness of xenophobic stereotypes in cultural artifacts and their role in nationalistic narratives~\citep{smith2002politics, leerssen2006nationalism, gordy2010culture, khandy2021pop} make it imperative to audit large AI models and datasets on those grounds.

\subsubsection{Risks}

Recent work on enumerating risks in large language models ~\citep{weidinger2021ethical, weidinger2022taxonomy}  categorized the potential downstream harms across six types: 1) discrimination, exclusion and toxicity; 2) information hazards; 3) misinformation harms; 4) malicious uses; 5) human-computer interaction harms; and 6) automation, access and environmental harms. These categories intersect with material, representational and wider societal harms through which we've been discussing the potential impact of xenophobia in sociotechnical systems, adding another layer of analysis. LLMs may potentially surface direct xenophobic discrimination, they may pose information hazard by revealing sensitive or private information about vulnerable individuals, spread misinformation that disparages out-groups, be incorporated in technologies weaponized by malicious xenophobic users, subtly reinforce discriminatory stereotypes via repeated human-computer interaction, and indirectly lead to disparate socioeconomic and environmental impact. Prior studies have identified a concerning aptitude of such systems towards accurately emulating extremist views when prompted~\citep{gpt3radicalise}, and numerous ethnic biases~\citep{li2020unqovering}. Perhaps unsurprisingly, these problems were shown to arise in multi-modal systems as well~\citep{multistereotype, cho2022dall, multimodaldataissues}. As this type of content may potentially influence and radicalise people into far-right ideologies at an unprecedented scale and pace, more focused investments in data curation are needed to address underlying risks~\citep{bender2021dangers}.

\subsubsection{Promise}

AI systems need to be designed to transcend ethno-nationalistic exclusionary cultural narratives and challenge entrenched xenophobic views. This would involve developing better ways of incorporating domain knowledge on the cross-cultural underpinnings of social friction, and breaking away from a stochastic re-enactment of historical patterns. This stands in stark contrast to the historical AI approach, centered around supervised learning and reconstructing pre-existing data. There is a pressing need to imbue general AI systems with capacities for \emph{understanding} and \emph{reasoning} about the world, and therefore the ability to integrate and reinterpret information, deriving less-biased and better reasoned conclusions about social phenomena.

Recent advances in understanding and mitigating social biases in language models~\citep{zmigrod2019counterfactual, huang2019reducing, li2020unqovering, liang2021towards, ousidhoum2021probing, nozza2021honest}, and computer vision systems~\citep{joo2020gender, wang2020revise}, coupled with a development of benchmarks~\citep{nangia2020crows, nadeem2020stereoset} for identification of malignant stereotypes in AI models, may pave the way towards quantifying the extent of xenophobic bias and harm in large pre-trained AI systems. Yet, for contemporary base models, the solution likely needs to be multi-faceted and case-specific. 

Multi-lingual~\citep{huang2019multi, xu2021layoutxlm, srinivasan2021wit} models may help incorporate a wider variety of viewpoints, especially if grounded in geospatially diverse imagery aimed at improving representation - though there are still open challenges when it comes to ensuring equal performance~\citep{wang2021assessing, choudhury2021linguistically, ramesh2023fairness}. Self-supervised learning in computer vision has shown promise in terms of reducing the amount of bias~\citep{goyal2022vision}, as well as presenting new ways to incorporate fairness objectives into model training~\citep{tsai2021conditional}. Data curation~\citep{scao2022bloom} plays a central role in mitigating the effects of historical bias. Mechanisms for incorporating comparative and natural language human feedback to refine and improve model outputs~\citep{llm_human_feedback, language_feedback} present another useful avenue, as do efforts for soliciting human preferences on conversational rules, and highlighting sources of information to improve factuality of claims~\citep{llm_sparrow}. This myriad of technical mitigations needs to be incorporated in a more holistic participatory approach, to avoid pitfalls of technosolutionism and empower marginalised communities in AI system design.
\section{Towards xenophilic AI systems}
\label{sec:xenophilic}

\subsection{A moral imperative}

Crucially, in terms of scope and impact, many of the AI tools and services discussed in this paper have global reach~\citep{crawford2021atlas}. They easily traverse national boundaries and – in the case of social media platforms – involve billions of users worldwide. Cumulatively, these technologies therefore have a powerful shaping effect on the contours of social relationships at a global level. Considered in this light, it is worth pausing to more fully assess the combined effect of the practices outlined in this paper. 

Taken together, xenophobic discrimination in the domain of AI can be best understood as a \emph{compound} or \emph{structural} phenomenon and set of experiences: people are not only exposed to automated xenophobic bias via interaction with one or two services, but across the entire range of digital services they interact with. Crucially, in an age of growing displacement and movement of people, a person who is new to a country, a non-native speaker, or a carrier of identity traits that attract xenophobic sentiment may expect to receive, among other things: (1) higher rates of harassment on social media, (2) problematic treatment by immigration services, (3) increased chance of medical error, (4) prejudicial access to employment opportunities, and (5) algorithmically embedded stereotypes, in addition to whatever other forms of disadvantage they may encounter. They are therefore in a position of \emph{intersectional vulnerability}~\citep{crenshaw2017intersectionality}: there is added precarity that comes with the notion of being “foreign” in a digital age. This is a uniquely dangerous position to occupy and one that we believe warrants special protection when it comes to the design and deployment of algorithmic systems.

Given this wider context, technologists face an important choice when it come to xenophobia and AI. They can build systems that heighten tensions between different groups and compound existing axes of disadvantage or they can promote a different, more inclusive, ideal – helping to mitigate the impact of xenophobia by learning from those affected and by taking concrete measures to forestall these effects. This aspiration, to build better systems that are capable of promoting civic inclusion and deescalating othering dynamics, we term “xenophilic technology”. When successful, xenophilic technology helps to build relationships between people of different backgrounds and to enable the rich and productive sharing of cultural differences.

Ultimately, the risks outlined in this paper indicate that the problem of xenophobic bias in AI systems cannot go unaddressed. Instead, we believe that technologists should partner with domain experts to jointly take the lead in this area, drawing upon sources such as the Universal Declaration on Human Rights (UDHR) which is quite explicit in its opposition to discrimination on the basis of national origin or related grounds~\citep{prabhakaran2022human}. On this point, Article 2 of the UDHR states that “everyone is entitled to all the rights and freedoms set forth in this Declaration without distinction of any kind, such as race, colour, sex, language, religion, political or other opinion, national or social origin, property, birth or other status.” Additionally, the global reach of technology also brings with it significant promise. By addressing the potential for material harm, representational harm, and risk to people's rights, responsibly designed technology may help bridge the divide between communities, minimize material and representational harms to foreigners and non-citizens, ensure that these people are secure in their standing as members of the relevant communities, and help to ensure that human rights are respected.

\subsection{Measurement and mitigation}

Evaluating and mitigating the potential xenophobic impact of AI systems is not common practice. In this section we highlight the technical considerations in relevant areas of ML fairness research that may prove useful for addressing this unmet need and help us move away from more hegemonic ML fairness approaches~\citep{Weinberg_2022}.

\subsubsection{Measuring xenophobia}

Most approaches to measuring fairness rely on an understanding of marginalised identities, coupled with the availability of the corresponding information in the underlying data. This is why it's especially concerning that there are deep issues with the overall quality of race and ethnicity data across observational databases~\citep{polubriaginof2019challenges}, which often contain insufficiently fine-grained and static racial and ethnic categories~\citep{strmic2018race, hanna2020towards}, involve discrepancies between recorded and self-reported identities~\citep{moscou2003validity} and come with data collection~\citep{varcoe2009harms, ford2005conceptualizing, routen2022strategies, andrus2022demographic}, privacy~\citep{ringelheim2011ethnic} and governance challenges. Despite recent integration efforts~\citep{muller2021linking}, better data~\citep{pushkarna2022data} and model~\citep{mitchell2019model, crisan2022interactive} documentation are required to promote the understanding of ethnic representation and data provenance.

Fairness approaches rooted in comparing model performance and impact across groups need to be more cognizant of the normative challenges that arise when defining group identities~\citep{leben2020normative}, and there also needs to be a deeper engagement with domain experts to establish consistent and meaningful ways of identifying in-groups and out-groups within the context of each AI application. The lack of frameworks and benchmarks for assessing xenophobic impact makes it especially hard to evaluate the impact of systems with a potentially global scope, which is a definitive concern. In the interim, practitioners should be encouraged to include ethnicity and nationality in routine ML fairness evaluation, in intersection with other relevant sensitive attributes~\citep{newinter}, and drive technical innovation towards the development of methods that can transcend the confines of pre-defined categorizations~\citep{pmlr-v139-jalal21b}.

Counterfactual~\citep{kusner2017counterfactual, pfohl2019counterfactual, joo2020gender} and contrastive~\citep{chakraborti2020contrastive} methodologies may prove useful for more concretely identifying the pathways of discrimination, while leveraging expert knowledge. Most such methods would still require a meaningful categorization of group identities, and may not be applicable in certain contexts where certain social categories may not lend themselves to meaningful counterfactual manipulation, making it impossible to reliably assess the truthfulness or counterfactuals~\citep{kasirzadeh2021use}. Individual fairness~\citep{dwork2012fairness, dwork2018individual, sharifi2019average, gupta2021individual} may prove applicable in absence of reliable categorizations, and there is ongoing research on eliciting~\citep{jung2019eliciting, bechavod2020metric} or learning~\citep{ilvento2019metric, mukherjee2020two} similarity measures for its practical application.

Quantitative measures of impact should be complemented by qualitative assessments~\citep{van2013psychometrically, krumpal2012estimating,  olonisakin2021xenophobia}, and comprehensive participatory approaches towards deeper and more representative assessments of xenophobic impact. Wider community participation~\citep{martin2020participatory, Bondi_2021, Birhane_2022} is key when creating benchmarks, soliciting feedback, and empowering the affected communities to inform policy and AI system design. As an ideological and social phenomenon, xenophobia may be targeted at an ever-evolving conception of what is foreign, according to local histories, cultural and political context~\citep{mamdani2018citizen, mamdani2012define}. Understanding who is deemed foreign, by whom, on what grounds, and in light of which set of unfolding circumstances, requires knowledge that is local, situated and contemporaneous.

Evaluation of xenophobic bias should also help re-center the question of discrimination onto the perpetrators. AI systems may be used with malicious intent, towards spreading hate and amplifying discrimination against vulnerable populations. Human bias may also skew the outcomes of assistive AI tech with human in the loop. To address the root cause of xenophobia, it is important to understand the role of AI systems in in-group radicalization, as well as how AI systems shift power. In terms of better understanding outcomes, we suggest for practitioners to consider, in line with what we propose in the paper, material harms, recognition harms and rights harms. This is merely a step towards a more holistic approach, as there is a need to also engage with the deeper questions of social justice~\citep{fairness_margins, schwobel2022long}.

\subsubsection{Mitigation strategies}

The complexity and pervasiveness of xenophobia in society and consequently in the data used to train AI systems calls for the development of both technical and non-technical mitigation strategies for preventing the adverse impact of such systems on the most vulnerable and marginalised groups in society, citizens and non-citizens alike. These are currently lacking, in part due to a systemic failure to recognize the importance of xenophobic harms, and in part due to already discussed deficiencies in recording and utilising the relevant sensitive information, compounded by insufficient investments in participatory approaches.

Until such comprehensive frameworks for identifying and quantifying xenophobic harms are fully developed, practitioners may want to consider alternative approaches, tailored for systems with missing or incomplete data, that are capable of providing partial safety and robustness guarantees without a full categorical specification. This may involve considering plausible groupings based on the available inputs, and improving worst-case or average-case performance across such plausible group structures~\citep{kearns2018preventing, kim2018fairness, hashimoto2018fairness}. The use of proxy targets~\citep{gupta2018proxy, grari2021fairness} is an alternative, though it comes with risks of stereotyping. Practitioners need to ensure robust evaluation~\citep{mandal2020ensuring} while accounting for uncertainty and incompleteness~\citep{awasthi2021evaluating} in the data. Purely technical approaches rarely map onto socially acceptable mitigation outcomes~\citep{slowik2021algorithmic}, making it imperative to avoid technosolutionism and to embrace interdisciplinary efforts.

Transparency and model audits~\citep{raji2020closing} need to complement such fairness assessments, especially given the difficulties involved with defining comprehensive quantitative measures, that would consistently capture all instances of xenophobic harms. Advances in AI explainability research~\citep{gilpin2018explaining, bhatt2020explainable} may be helpful in such audits~\citep{abdul2018trends, hall2019proposed, bibal2021legal}, enabling a rapid identification of model harms~\citep{begley2020explainability} possibly enriched with counterfactual explanations~\citep{sokol2019counterfactual, verma2020counterfactual}. Explainability methods may also be helpful in identifying vulnerable groups~\citep{strumke2022explainability} within each AI application context, and provide a fine-grained breakdown of model performance~\citep{sharma2020certifai}. These efforts need to be coupled with the appropriate escalation procedures and enable recourse -- which may require a greater focus on designing appropriate interventions~\citep{action_recourse}.
\section{Conclusions}

Despite being one of the key drivers of discrimination and conflict worldwide, xenophobic bias is yet to be formally recognised and comprehensively assessed in AI system development. This blind spot introduces a considerable risk of amplifying harms to marginalised communities and fuelling fear and intolerance towards foreigners, which may lead to violence and conflict both at the societal level and more widely.

Our overview of several prominent technological use cases -- social media, immigration, healthcare, employment and the advances in the development of foundation models -- reveals how historical and contemporary xenophobic attitudes may manifest in data and AI systems. Moreover, we illustrate the effect of these tendencies by drawing upon three categories of potential xenophobic harms: those that cause direct material disadvantage to people perceived as foreign, representational harms that undermine their social standing or status, and the wider societal harms which include barriers to the successful exercise of civic and human rights. Significantly, given the scope and depth of AI services, people who experience xenophobic bias in one domain of automation can also expect to experience unequal treatment across a range of other domains, further compounding their situation.

In this context, we suggest that the development of xenophilic technology is crucial for the ethical design and deployment of AI systems, understood as systems that treat people equitable at the societal level. AI should help promote civic inclusion, oppose the malignant “othering” of marginalised groups, and be centered around the potential to cultivate the kind of rich and productive cross-cultural discourse that is appropriate for a world like our own. Coordinated interdisciplinary action, participatory frameworks, and an unwavering commitment of technologists are required to fashion such a future, along with technical innovation in addressing the existing technological bases of xenophobia in ML systems. Xenophilic technology should be understood as merely a part of a much wider set of solutions, within and across societies, avoiding the pitfalls of technosolutionism.

\begin{acks}
We would like to thank Stevie Bergman, William Isaac and Shakir Mohamed for the support and insightful feedback that they provided for this paper.

Any opinions presented in this paper represent the personal views of the authors and do not necessarily reflect the official policies or positions of their organisations.
\end{acks}

\bibliographystyle{ACM-Reference-Format}
\bibliography{main}


\end{document}